\documentclass[12pt]{iopart}
\usepackage{iopams,graphicx,cite}

\newcommand{\expp}[1]{ \mathop\mathit{e}\nolimits^{#1}}

\newcommand{\ud}[2][]{\mathrm{d}^{#1}{#2}\,}

\newcommand{\vd}[2][]{\mathrm{d}^{#1}{#2}}

\newcommand{\ie}{\mbox{i.e.}}

\renewcommand{\Re}{\mathop\mathrm{Re}\nolimits}
\renewcommand{\Im}{\mathop\mathrm{Im}\nolimits}

\newcommand{\GR}{G_\mathrm{R}}
\newcommand{\SigmaR}{\Sigma_\mathrm R}
\newcommand{\vect}{\mathbf}

\newcommand{\phim}{\phi_m}
\newcommand{\phiM}{\phi_M}
\newcommand{\dm}{\mathit{\mathrm{\Delta}m}}

\newcommand{\eqref}[1]{(\ref{#1})}

\begin{document}

\title{Particle propagation in cosmological backgrounds}

\author{Daniel Arteaga}
\address{Departament de F\'\i sica Fonamental. Facultat de F\'\i sica. Av. Diagonal 647, 08028 Barcelona (Spain)}
\ead{darteaga@ub.edu}

\begin{abstract}
We study the quantum propagation of particles in cosmological backgrounds, by considering a  
doublet of massive scalar fields propagating in an 
expanding universe, possibly filled with radiation. We focus on the dissipative effects
related to the expansion rate. At first order,
we recover the expected result 
that the decay rate is determined by the local temperature.  Beyond linear order, the decay rate has an additional contribution governed by the expansion parameter. This latter contribution is present even for stable particles in the vacuum.  Finally, we analyze the long time behaviour 
of the propagator and briefly discuss 
applications to the trans-Planckian question.
\end{abstract}

\section{Introduction}

In this contribution we study the quantum propagation of 
particles in a cosmological background. We are particularly interested in 
understanding 
the dissipative phenomena  
related to the time
dependence of the 
metric. 
To this end, we analyze the propagator of a massive particle
which interacts with a massless radiation field 
in an expanding universe. Several points must be considered.

First, 
we are dealing with an interacting field theory in a curved spacetime. 
In this situation,
the asymptotic \emph{in} and \emph{out} vacua
generally do not coincide. 
Being interested in expectation values, 
rather than in
\emph{in-out} matrix elements,
we adopt
the Keldysh-Schwinger formalism, or Closed Time Path 
(CTP) method \cite{Schwinger61,Keldysh65,ChouEtAl85} in curved spacetime \cite{Jordan86,CalzettaHu87,CamposVerdaguer94,Weinberg05}.

Second, as it is well known, 
in a curved spacetime there is no 
single definition for the vacuum nor for the concept of particle. 
We 
face this issue by working within 
the adiabatic approximation \cite{BirrellDavies}:
the massive particles 
will have
their Compton wavelengths much smaller than the typical curvature radius of 
the universe (in our case, the Hubble radius). 

Third, as explained in \cite{ArteagaParentaniVerdaguer04a}, in 
theories such as QED or perturbative quantum gravity, dissipative effects 
appear 
only
at two loops, 
because the one-loop 
diagrams which  
could 
lead to dissipation 
vanish on the mass shell.
Here, in order 
to keep the calculations simple, 
we have  chosen a simple, yet physically meaningful, model 
which exhibits dissipation at one loop. We expect  the behaviour of 
QED or perturbative quantum gravity to be similar at two loops.

In this contribution we 
compute the retarded self-energy of the lightest 
field in a massive doublet
which propagates
in a thermal bath of massless particles in an expanding universe, and from it we extract the decay rate. Notice that in the Minkowski vacuum the excitations of the lighter field are stable, hence the decay rate is zero.
We concentrate on the physical insights and summarize the main results. A more detailed account will be given in  separate publications \cite{ArteagaParentaniVerdaguer06,ArteagaParentaniVerdaguerInPrep}

The contribution is organized as follows. In section 2 we 
introduce the 
model and motivate the use of the adiabatic approximation for the massive fields. 
In section 3  we present the results for the imaginary part of the self-energy and the decay rate. In section 4 we study the time evolution of the interacting propagator. Finally, in section 5 we summarize 
the main points of the contribution and discuss its relevance to the trans-Planckian question. We use a system of natural units 
with $\hbar=c=1$, and the metric has the signature $(-,+,+,+)$. 

\section{The model}

We 
consider 
spatially isotropic and homogeneous
Friedmann-Lema\^itre-Robertson-Walker models with flat spatial sections:
\begin{equation}
	\vd s^2 = - \vd t^2 + a^2(t) \vd {\vect x}^2.
\end{equation}
The particle model is the following:
two massive fields $\phim$, 
and $\phiM$, interacting with a massless field, $\chi$, via a trilinear coupling. 
The total action is $S = S_m + S_M + S_\chi + S_\mathrm{int}$,
where each term is given by
\numparts
\begin{eqnarray}
	S_m = \frac{1}{2} \int \ud t \ud[3] {\vect x} a^3(t)  \left( (\partial_t \phim)^2 - \frac1{a^2(t)} (\partial_\vect x \phim)^2 - m^2 \phim^2 \right),\\
	S_M = \frac{1}{2} \int \ud t \ud[3] {\vect x} a^3(t)  \left( (\partial_t \phiM)^2 - \frac1{a^2(t)} (\partial_\vect x \phiM)^2 -  M^2 \phiM^2 \right),\\
	S_\chi = \frac{1}{2} \int \ud t \ud[3]{\vect x} a^3(t) 
	\left(  (\partial_t \chi)^2 - \frac1{a^2(t)} 
	(\partial_\vect x \chi)^2 - \xi R(t) 
	\chi^2 \right) ,\\
	S_\mathrm{int} = g M \int \ud t \ud[3]{\vect x} a^3(t)  \phim \phiM \chi,
\end{eqnarray}
\endnumparts
with  $R(t)$ being the Ricci scalar. We 
assume that the massless field is conformally coupled to gravity, so that $\xi = 1/6$.
It is 
useful to 
work with rescaled massive fields 
defined by $\bar\phi (t,\vect x):= [-g(t,\vect x)]^{1/4}\phi(t,\vect x) = 
a^{3/2}(t) \phi(t,\vect x)$. 

We consider the two massive fields having large masses but with a small mass difference $\dm := M-m \ll M $. 
As shown in \cite{ArteagaParentaniVerdaguer05}, the model can be 
interpreted as a 
field-theory description of a 
relativistic
two-level atom 
(of mass $m$ and energy gap $\dm$)
interacting with a 
scalar radiation field $\chi$. 
The radiation field $\chi$ is assumed to be at some conformal temperature $\theta$ (which can eventually be zero). The corresponding physical temperature, as well as the Hubble rate $H(t) := \dot a(t)/a(t)$, are chosen to be 
much smaller than the masses of the fields.
These restrictions ensure that
the number of massive particles is strictly conserved.
The non-trivial dynamics concerns the transitions between the two massive 
fields accompanied by emission and absorption of massless quanta.

In a curved spacetimes it is not a trivial task
to compute even the free field vacuum 
propagators. 
For massless conformally coupled fields there is a natural vacuum state,
the conformal vacuum. Propagators in this vacuum, when expressed in conformal time, essentially correspond to the flat spacetime propagators.
For the
massive 
fields, 
rather than attempting to find the exact free
propagator, we will exploit 
the fact that their Compton wavelengths 
is much smaller than 
the Hubble 
length $H^{-1}$.  
In this regime, the adiabatic (WKB) 
approximation is valid and explicit expressions for the free propagators can be computed \cite{BirrellDavies,ArteagaParentaniVerdaguer06} ---see for instance \eqref{FreeRetarded}.

\section{The self-energy and decay rates}

In this section we consider the 
interacting retarded Green function 
\begin{equation}
G_\mathrm{R}(t,t';\vect p):=\theta(t-t')
\langle{[\hat\phi_{m\vect p}(t),\hat\phi_{m\vect p}(t')]}\rangle 
\end{equation}
within the adiabatic approximation.
It is 
related to
the retarded self-energy $\SigmaR$ via \cite{ArteagaParentaniVerdaguer05}
\begin{eqnarray}
	\fl \GR(t,t';\vect p) = \GR^{(0)}(t,t';\vect p)\nonumber \\  -i \int \ud s \ud {s'} \sqrt{-g(s)} \sqrt{-g(s')} \GR^{(0)}(t,s;\vect p) \SigmaR(s,s';\vect p) \GR(s',t';\vect p)
\end{eqnarray}
where $\GR^{(0)}(t,t';\vect p)$ is the free retarded propagator. 
In terms of the rescaled 
fields, 
$\bar \phi(t; \vect p) = a^{3/2}  \phi(t; \vect p)$,
the above relation becomes
\begin{equation}\label{Dyson}
\fl
	\bar \GR(t,t';\vect p) = \bar\GR^{(0)}(t,t';\vect p) -i 
	\int \ud s \ud {s'}  \bar\GR^{(0)}(t,s;\vect p) 
	\bar\SigmaR(s,s';\vect p) \bar\GR(s',t';\vect p).
\end{equation}

We 
assume that the massive fields are in the adiabatic vacuum, 
and that the 
massless field $\chi$ is in a 
thermal state, 
characterized by a
fixed conformal temperature $\theta$. 
We will compute the imaginary part of the one-loop self energy to order $g^2$ 
in the 
adiabatic approximation, evaluated at the mass shell. 
It will be evaluated in 
a a frequency representation around the average time coordinate $T=(t_1+t_2)/2$, by Fourier-transforming with respect to the difference coordinate $\Delta=t_1-t_2$, which amounts to a local frequency representation (it is further analyzed in next section). As for the spatial part, we 
work in the momentum 
representation 
to exploit conservation of the conformal momentum.

\subsection{Linear approximation to the scale factor}

As a first step, we approximate the evolution of the scale factor by a linear expansion:
\begin{equation}
	a(t)\approx a(T)[1+H(T)(t-T)].
	\label{firstord}
\end{equation}
This approximation for the scale factor is appropriate when 
considering physical temperatures which are much larger 
than the expansion rate 
(but still much smaller than the 
fields masses). On the mass shell, thermal corrections will dominate over curvature corrections, since the thermal energy scale is much larger than the curvature energy scale. Therefore, we expect the on-shell self-energy to be governed by the thermal bath at the instantaneous physical temperature at each moment of the expansion, $\theta/a(T)$.

The explicit calculation\cite{ArteagaParentaniVerdaguer06} confirms that the imaginary part of the on-shell self-energy is given by that of a thermal bath in Minkowski at a physical temperature $\theta/a(T)$.  In the limit in which the atoms are at rest this result is\cite{ArteagaParentaniVerdaguer05,ArteagaParentaniVerdaguer06}
\begin{equation}
	\Im \bar\SigmaR (m,T;\vect 0) = -  \frac{g^2}{8\pi} M \dm \, 
	n_{\theta/a(T)} (\dm),
\end{equation}
where $n_{\theta/a(T)} (\dm)$ is the Bose-Einstein function:
\begin{equation}
	n_{\theta/a(T)} (\dm) := \frac{1}{\expp{\dm\, a(T)/\theta}-1}.
\end{equation}
As in Minkowski spacetime, the self-energy corresponds to a decay rate,
\begin{equation}
	\Gamma = -\frac{1}{m} \Im \bar\SigmaR (m,T;\vect 0) = \frac{g^2}{8\pi} \dm \, 
	n_{\theta/a(T)} (\dm),
\end{equation}
which amounts to the probability per unit time for the lightest state to absorb a massless particle from the thermal bath.

\subsection{Beyond linear order: vacuum effects}

When the expansion rate of the universe is of the order of the temperature or larger, vacuum effects become relevant. Energy conservation does not hold for energy scales of the order of the expansion rate, and therefore we expect new channels for the particle decay which will contribute to the imaginary part of the self-energy.

In order to study the vacuum effects we need to choose a explicit model for the evolution of the scale factor. For instance, in the case of de Sitter,
\begin{equation}
	a(t)=a(T)\expp{H(t-T)},
\end{equation}
the vacuum contribution to the imaginary part of the retarded self-energy given by \cite{ArteagaParentaniVerdaguerInPrep}
\begin{equation}
	\Im \bar\SigmaR (m,T;\vect 0) = -  \frac{g^2}{8\pi} M \dm \, 
	n_{H/(2\pi)} (\dm),
	\end{equation}
which coicides with the self-energy in a Minkowski thermal bath at a temperature $H/(2\pi)$. The result is not unexpected since the effective de Sitter temperature \cite{Mottola85} is recovered. The corresponding decay rate
\begin{equation}
	\Gamma = -\frac{1}{m} \Im \bar\SigmaR (m,T;\vect 0) = \frac{g^2}{8\pi} \dm \, 
	n_{H/(2\pi)} (\dm),
\end{equation}
amounts for the probability per unit time for the lightest field to emit a massless particle. Energy conservation forbids this process in Minkowski spacetime, but this restriction does not apply in an expanding universe.

\section{Retarded propagator and self-energy in cosmology}

In expanding universes the propagators are no
longer time-translation invariant. 
We can nevertheless always 
express the propagator 
in a frequency representation,
\begin{equation} \label{mixed}
	\bar\GR(\omega,T;\vect p):=\int \ud \Delta \expp{i\omega\Delta} 
	\bar\GR(T+\Delta/2,T-\Delta/2;\vect p) \, .
\end{equation}
For short time differences as compared to the inverse expansion rate, \ie, $|t-t'|\ll H^{-1}$, \eqref{Dyson} can be diagonalized:\begin{equation} \label{shortime}
	\bar\GR(\omega,T;\vect p) = \frac{ -i}
	{[-i \bar G^{(0)}(\omega,T;\vect p)]^{-1} + 
	\bar\SigmaR(\omega,T;\vect p)  }\, .
\end{equation}
Fourier-transforming again we get the short-time behavior:
\begin{equation}\label{shortimetime}
	\bar G_\mathrm{R}(t,t';\vect p) = \frac{-i}{ R_\vect p(T)} \sin\left[R_\vect p(T)(t-t')\right] \expp{-\Gamma_\vect p(T)(t-t')/2} \theta(t-t').
\end{equation}
with
\begin{equation} \label{R}
	\fl R^2_\vect p(T) :=  E^2_{\vect p}(T) + \Re\bar\SigmaR(E_\vect p,T;\vect p) := m^2 + \frac{\vect p^2}{a^2(T)} + \Re\bar\SigmaR(E_\vect p,T;\vect p)
\end{equation}
and 
\begin{equation} \label{gamma}
	\Gamma_\vect p(T) := -\frac{1}{R_\vect p(T)} \Im\bar\SigmaR(E_\vect p,T;\vect p).
\end{equation}
Therefore one recovers the usual interpretation, in which the real part 
of the self-energy corresponds to the energy shift, and  
in which the 
imaginary part corresponds to the decay rate. Notice that both quantities depend in general
 on time. 

One may also be interested in considering large time lapses, and in this case the frequency representation of the propagator around the average time does not make sense. Lifting the short-time requirement, and only imposing the adiabatic approximation, the following expression for the evolution of the retarded propagator is found \cite{ArteagaParentaniVerdaguer06}:
\begin{eqnarray}\label{AdiabaticLarge}
	\fl\bar G_\mathrm{R}(t_1,t_2;\vect p) = \frac{-i}{  \sqrt{ R_\vect p(t_1) R_\vect p(t_2)}} \sin\left({\int^{t_1}_{t_2} \ud{t'} R_\vect p(t')} \right) \expp{-{\int^{t_1}_{t_2} \ud{t'} \Gamma_k(t')/2} }  \theta(t_1-t_2).
\end{eqnarray}
Notice that the long-time evolution of the propagator can be expressed in terms of integrals of quantities evaluated in the local frequency representation.  Two time scales are clearely separated: the interaction timescale, in which the interaction process take place and in which the self-energy is evaluated, and the evolution timescale, which can be much longer and during which the propagators deviate significantly from the corresponding Minkowski expression.
Equation (\ref{AdiabaticLarge}) can be derived in a very similar way as the well-known adiabatic approximation for the free retarded propagator \cite{BirrellDavies}:
\begin{equation}\label{FreeRetarded}
	\bar G^{(0)}_\mathrm{R}(t_1,t_2;\vect p) = \frac{-i}{  \sqrt{ E_\vect p(t_1) E_\vect p(t_2)}} \sin\left({\int^{t_1}_{t_2} \ud{t'} E_\vect p(t')} \right) \theta(t_1-t_2).
\end{equation}

\section{Summary and discussion}

The goal of this contribution is to analyze the quantum effects in the propagation 
of 
interacting fields in a cosmological background. 
This issue 
may play an important role in justifying the non-trivial dispersion relations which have been used when addressing the trans-Planckian question in the context
of black holes \cite{Unruh81,Jacobson91,Unruh95,BalbinotEtAl06}
and cosmology \cite{MartinBrandenberger01,MartinBrandenberger03,Niemeyer00,NiemeyerParentani01}.
Interactions could indeed
significantly modify 
the field propagation
when approaching the event horizon of a black hole \cite{BarrabesFrolovParentani00,Parentani01,Parentani02,Parentani02b}
or at primordial stages of inflation \cite{ArteagaParentaniVerdaguer04a}.

In our model, the masses of the fields were assumed to be much larger than the expansion rate of the universe. This was a key assumption, because it allowed to introduce the adiabatic (WKB) approximation, which not only makes the problem solvable, but also allows having a well-defined particle concept even in absence of asymptotic regimes. Within this approximation, the time-evolution of the interacting propagators can be computed from the integral of the retarded self-energy, evaluated on-shell in a frequency representation around the mid time.

The imaginary part of the self-energy determines the decay of the retarded propagator, and hence it is an expression of the dissipative properties. For temperatures higher than the expansion parameter the decay of the propagator is determined by the local temperature at each moment of expansion. For lower temperatures, the decay of the propagator is driven by the expansion rate of the universe. This second contribution, which is present even in the vacuum, can be interpreted as being a consequence of the absence of energy conservation at those energy scales comparable to the expansion rate.

The decay rate, derived from the imaginary part of 
the self-energy, has a secular character. Even small decay rates 
could thus give an important effect when integrated over large periods of time.  The exact significance of the generically dissipative properties of the propagator will be further analyzed elsewhere \cite{ArteagaParentaniVerdaguerInPrep}.

\ack

I am very grateful with Renaud Parentani and Enric Verdaguer for a critical reading of the manuscript.  This work is partially supported by the Research Projects MEC FPA2004-04582-C02-02 and DURSI 2005SGR-00082.


\section*{References}


\begin{thebibliography}{10}

\bibitem{Schwinger61}
J.~S. Schwinger.
\newblock {\em J. Math. Phys.}, 2:407, 1961.

\bibitem{Keldysh65}
L.~V. Keldysh.
\newblock {\em Zh. Eksp. Teor. Fiz}, 47:1515, 1965.
\newblock [Sov. Phys. JEPT 20:1018, 1965].

\bibitem{ChouEtAl85}
K.-C. Chou, Z.-B. Su, B.-L. Hao, and L. Yu.
\newblock {\em Phys. Rept.}, 118:1--131, 1985.

\bibitem{Jordan86}
R.~D. Jordan.
\newblock {\em Phys. Rev. D}, 33:444--454, 1986.

\bibitem{CalzettaHu87}
E.~Calzetta and B.~L. Hu.
\newblock {\em Phys. Rev. D}, 35:495--509, 1987.

\bibitem{CamposVerdaguer94}
A. Campos and E. Verdaguer.
\newblock {\em Phys. Rev. D}, 49:1861--1880, 1994.

\bibitem{Weinberg05}
S. Weinberg.
\newblock {\em Phys. Rev. D}, 72:043514, 2005.

\bibitem{BirrellDavies}
N.~D. Birrell and P.~C.~W. Davies.
\newblock {\em Quantum fields in curved space}.
\newblock Cambridge University Press, Cambridge, England, 1982.

\bibitem{ArteagaParentaniVerdaguer04a}
D. Arteaga, R. Parentani, and E. Verdaguer.
\newblock {\em Phys. Rev. D}, 70:044019, 2004.

\bibitem{ArteagaParentaniVerdaguer06}
D. Arteaga, R. Parentani, and E. Verdaguer.
\newblock To appear in {\em Int. J. Theor. Phys.}

\bibitem{ArteagaParentaniVerdaguerInPrep}
D. Arteaga, R. Parentani, and E. Verdaguer.
\newblock In preparation.

\bibitem{ArteagaParentaniVerdaguer05}
D. Arteaga, R. Parentani, and E. Verdaguer.
\newblock {\em Int. J. Theor. Phys.}, 44:1665--1689,
  2005.
  
\bibitem{Mottola85} E. Mottola, {\em Phys. Rev. D}, 31:754--766, 1985.


\bibitem{Unruh81}
W.~G. Unruh.
\newblock {\em Phys. Rev. Lett.}, 46:1351--1353, 1981.

\bibitem{Jacobson91}
T. Jacobson.
\newblock {\em Phys. Rev. D}, 44:1731--1739, 1991.

\bibitem{Unruh95}
W.~G. Unruh.
\newblock {\em Phys. Rev. D}, 51:2827--2838, 1995.

\bibitem{BalbinotEtAl06}
R.~Balbinot, A.~Fabbri, S.~Fagnocchi, and R.~Parentani.
\newblock {\em Riv. Nuovo Cim.} 28:1--55, 2005 (gr-qc/0601079).

\bibitem{MartinBrandenberger01}
J. Martin and R. Brandenberger.
\newblock {\em Phys. Rev. D}, 63:123501, 2001.

\bibitem{MartinBrandenberger03}
J. Martin and R. Brandenberger.
\newblock {\em Phys. Rev. D}, 68:063513, 2003.

\bibitem{Niemeyer00}
J.~C. Niemeyer.
\newblock {\em Phys. Rev. D}, 63:123502, 2001.

\bibitem{NiemeyerParentani01}
J.~C. Niemeyer and R. Parentani.
\newblock {\em Phys. Rev. D}, 64:101301, 2001.

\bibitem{BarrabesFrolovParentani00}
C.~Barrab\`es, V.~Frolov, and R.~Parentani.
\newblock {\em Phys. Rev. D}, 62:044020, 2000.

\bibitem{Parentani01}
R.~Parentani.
\newblock {\em Phys. Rev. D}, 63:041503, 2001.

\bibitem{Parentani02}
R.~Parentani.
\newblock {\em Int. J. Theor. Phys.}, 41:2175--2200, 2002.

\bibitem{Parentani02b}
R. Parentani.
\newblock {\em Int. J. Mod. Phys.}, A17:2721--2726, 2002.




\end{thebibliography}
\end{document}